\documentclass[letterpaper, 10 pt, conference]{ieeeconf}  
\IEEEoverridecommandlockouts                              
\usepackage{epsfig} 
\usepackage{graphicx}
\usepackage{amsmath}
\usepackage{array}
\usepackage{amssymb}
\usepackage{verbatim}
\usepackage{cleveref}
\usepackage{color}
\usepackage{relsize}
\usepackage{tikz}
\usepackage{fancyhdr}
\usepackage{graphics}
\usepackage{setspace}
\usepackage{epstopdf}
\usepackage{multirow}
\usepackage{lipsum}
\usepackage{textcomp}
\usepackage{mathtools}
\usepackage{cite}
\usepackage{setspace}
\usepackage{booktabs}
\usepackage{breqn}
\usepackage{algorithm}
\usepackage[noend]{algpseudocode}
\usepackage{flushend}
\usepackage{multicol}
\usepackage{float}
\usepackage{mathtools,bm}
\DeclareMathOperator*{\argmin}{min}
\usepackage[noend]{algpseudocode}
\usepackage{booktabs}
\usepackage{dblfloatfix}
\usepackage{mathrsfs}
\usepackage{mathtools}

\fancypagestyle{firstpage}{%
  \lhead{\vspace{-0.75 cm} \small \textbf{{\fontfamily{cmss}\selectfont
2019 Annual American Control Conference (ACC)\\July 10–-12, 2019, Philadelphia, PA, USA}}}
}

\title{\large \bf
Sequential Optimization of Speed, Thermal Load, and Power Split in Connected HEVs$^*$}

\author{Mohammad Reza Amini$^{1}$, Xun Gong$^{1}$, Yiheng Feng$^{2}$, Hao Wang$^{1}$, Ilya Kolmanovsky$^{3}$, and Jing Sun$^{1}$
\thanks{*This paper is based upon the work supported by the United States Department of Energy (DOE) under award No. DE-AR0000797.}
\thanks{$^{1}$M. R. Amini, X. Gong, H. Wang, and J. Sun are with the Department of Naval Architecture \& Marine Engineering, University of Michigan, Ann Arbor, MI 48109 USA. Emails: {\tt\small \{mamini,gongxun,autowang,jingsun\}@umich.edu}}%
\thanks{$^{2}$Y. Feng is with the Transportation Research Institute, University of Michigan, Ann Arbor, MI, 48109, USA. Email: {\tt\small yhfeng@umich.edu}
}
\thanks{$^{3}$I. Kolmanovsky is with  the Department of Aerospace Engineering, University of Michigan, Ann Arbor, MI 48109 USA. Email: {\tt\small ilya@umich.edu}}%
}

\renewcommand{\baselinestretch}{1}
\begin{document}

\maketitle
\thispagestyle{firstpage}


\begin{abstract}
The emergence of connected and automated vehicles (CAVs) provides an unprecedented opportunity to capitalize on these technologies well beyond their original designed intents. While abundant evidence has been accumulated showing substantial fuel economy improvement benefits achieved through advanced powertrain control, the implications of the CAV operation on power and thermal management have not been fully investigated. In this paper, in order to explore the opportunities for the coordination between the onboard thermal management and the power split control, we present a sequential optimization solution for eco-driving speed trajectory planning, air conditioning (A/C) thermal load planning (eco-cooling), and powertrain control in hybrid electric CAVs to evaluate the individual as well as the collective energy savings through proactive usage of traffic data for vehicle speed prediction. Simulation results over a real-world driving cycle show that compared to a baseline non-CAV, 11.9\%, 14.2\%, and 18.8\% energy savings can be accumulated sequentially through speed, thermal load, and power split optimizations, respectively.
 \end{abstract}

\vspace{-0.15cm}
\section{INTRODUCTION}
\vspace{-0.15cm}
With the advent of connected and automated vehicles (CAVs) equipped with advanced sensors for perception and localization, communications and feedback have become more integrated and ubiquitous, leading to tangible benefits with enhanced safety and improved mobility\cite{sciarretta2015optimal}. Additionally, connectivity and autonomous driving technologies open up new dimensions for control and optimization of vehicle dynamics and powertrain systems. Over the past years, extensive studies have been carried out on powertrain optimization for electrified vehicles. Most of the recent CAV-related research, such as eco-driving and platooning, has been focused on reducing traction power related losses (see~\cite{guanetti2018control} and the references therein), while little has been reported on the use of external information (V2X) for coordinated power and thermal management (PTM) for the CAVs. 

In CAVs driving environment, by utilizing information from traffic systems and other surrounding vehicles, the vehicle speed can be adjusted to generate trajectories that avoid frequent acceleration and braking, thereby reducing vehicle fuel consumption and emissions~\cite{Zhen2018Traj,prakash2016assessing}. These types of driving behaviors that lead to energy saving are often referred to as eco-driving~\cite{barth2009energy}. The energy saving potential of eco-driving can be further enhanced by co-optimization of vehicle speed trajectory with vehicle powertrain control system operation. In addition to the powertrain system, efficient thermal management is a significant factor in the overall vehicle energy consumption optimization for electrified vehicles~\cite{AminiCDC18}. Thus, coordinated control of vehicle speed and the PTM system can potentially maximize the energy savings in electrified vehicles, including pure electric vehicles (EVs) and hybrid electric vehicles (HEVs). 

This paper proposes an energy optimization framework for HEVs by coordinating the speed trajectory planning with PTM system via a sequential optimization scheme with three optimization stages: (i) optimal speed trajectory planning using traffic information (i.e., eco-driving), (ii) thermal load optimization by exploiting the speed sensitivity of the air conditioning (A/C) system (i.e., eco-cooling), and (iii) powertrain optimization seeking optimal power split between the engine and the battery. An eco-driving model from our previous work~\cite{Zhen2018Traj} for connected HEVs in a congested urban traffic environment is considered. The vehicle queuing process is explicitly modeled by the shockwave profile model with consideration of vehicle deceleration and acceleration to provide a green window for eco-vehicle speed trajectory planning. For light-duty HEVs with relatively small batteries, air conditioning of the passenger compartment is the primary thermal load in summer, which represents the most significant auxiliary load on the electric battery~\cite{Rugh2008}. It has been found that the energy efficiency of the A/C system is dependent on vehicle speed~\cite{hwang2018}. In order to exploit this sensitivity for eco-cooling, we adopt a control-oriented thermal model of the A/C system from our previous work~\cite{hwang2018}, and design an energy-conscious hierarchical model predictive controller (H-MPC) for the A/C system with respect to the planned optimal speed trajectory from the first stage. Next, the optimized traction and thermal loads are used to determine the optimal power split ratio between the mechanical (engine) and electrical (battery) propulsion systems of the HEV using Dynamic Programming (DP). To this end, a simplified model of the HEV powertrain based on the power balance is developed and experimentally validated. The model is used to perform power split optimization, and evaluate the overall vehicle energy consumption using data collected from a fourth-generation Toyota Prius HEV.

The main contribution of this paper is the development of an optimization framework to assess the benefits of proactively using traffic information, which involves optimizing the vehicle speed, then optimizing the A/C thermal load accordingly, and eventually finding the optimal power split for an HEV operating in a CAV driving environment. The results of the developed sequential optimization help to understand the contribution of each optimization stage towards the overall vehicle performance and fuel economy improvements, based on which the effective structure of the integrated and real-time PTM controller can be determined. 

%

\vspace{-0.15cm} 
\section{Traffic Modeling}\vspace{-0.1cm} 
Surrounding traffic plays an important role in determining vehicle speed trajectories and therefore fuel consumptions, especially when the vehicle is approaching a signalized intersection. Through V2I communication, the infrastructure is able to receive real-time vehicle information. In combination with the traffic signal information, the infrastructure can predict vehicle queuing dynamics over a short horizon into the future, which allows for eco-driving trajectory planning.

\begin{figure}[h!]
  	\begin{center}
  		\includegraphics[width=0.85\columnwidth]{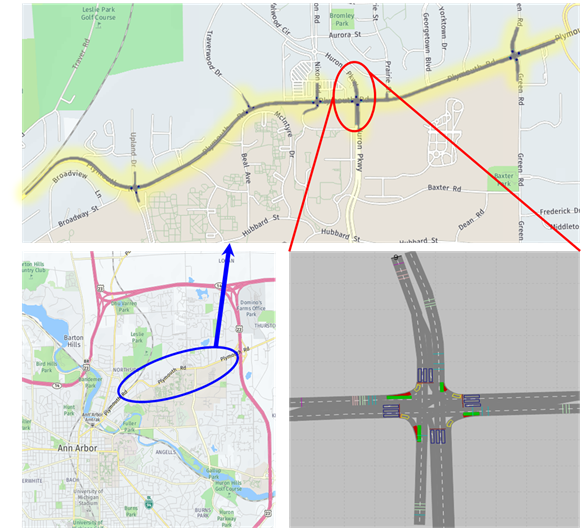} \vspace{-0.25cm}   
  		\caption{The corridor in Ann Arbor, MI for traffic modeling and simulation.}\vspace{-0.60cm} 
  		\label{fig:AnnArbor_CAVs_Model} 
  	\end{center}
  \end{figure}

To illustrate the approach, a six intersection corridor at Plymouth Rd, Ann Arbor is modeled. The stretch of the road represented in the simulations is about 2.2 miles and has two lanes for each direction. This stretch is one of the busiest commuting routes, serving US23 to the North Campus of the University of Michigan and downtown Ann Arbor. A microscopic traffic simulation software VISSIM~\cite{ptv2016ptv} is used to build the road network and simulate background traffic. To calibrate the simulation model to represent a congested traffic condition, real-world data were collected during PM rush hour (4:00PM-5:00PM) driving, including traffic volume, turning ratio, and traffic signal timing at each intersection with sampling time of $0.1~sec$. Fig.~\ref{fig:AnnArbor_CAVs_Model} shows the arterial corridor and the corresponding VISSIM model.\vspace{-0.25cm} 
 
\begin{figure}[h!]
  	\begin{center}
  		\includegraphics[width=0.8\columnwidth]{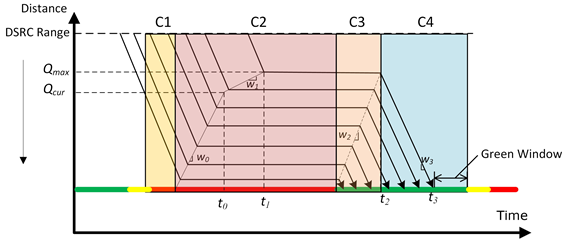} \vspace{-0.4cm}   
  		\caption{Queue profile prediction with shockwave profile model.}\vspace{-0.55cm} 
  		\label{fig:Queue_Prediction_SPM} 
  	\end{center}
  \end{figure}

The queuing process is explicitly modeled by the shockwave profile model (SPM)~\cite{wu2011shockwave} to provide a green window for eco-driving trajectory planning. The green window is defined as the time interval during which an eco-driving vehicle can pass through the intersection. The original SPM can only estimate the queue after the vehicles passed through the intersection. However, in this study we need to predict the queuing dynamics before the eco-driving vehicle arrival at the intersection. As a result, we modify the SPM to consider vehicle acceleration and deceleration process and make predictions of queuing dynamics. The entire queuing process within a signal cycle is shown in Fig.~\ref{fig:Queue_Prediction_SPM}. Four critical time instants are defined as (i) $t_0$: current time when the prediction is made, (ii) $t_1$: predicted time instant at which the maximum queue length $Q_{max}$ is reached (stop time of the front vehicle), (iii) $t_2$: predicted time instant at which the end of the queue starts to move (launch time of the front vehicle), and (iv) $t_3$: predicted time instant at which the end of the queue reaches the intersection (departure time of the front vehicle). This is also the start time of the green window. Note that the front vehicle is defined as the next downstream vehicle of the eco-driving vehicle in the same lane. 
\vspace{-0.37cm} 
 
The primary purpose of the queuing profile prediction is to determine the start of the green window ($t_3$). Four different shockwaves are identified in Fig.~\ref{fig:Queue_Prediction_SPM} to calculate $t_3$ step by step as (i) $w_0$: queuing shockwave speed until current time, (ii) $w_1$: predicted queuing shockwave speed until the maximum queue is reached, (iii) $w_2$: discharge shockwave speed, and (iv) $w_3$: departure shockwave speed. The time instant $t_3$ is when the departure shockwave $w_3$ arrives at the intersection. We assume that all vehicles are connected and broadcasting Basic Safety Messages (BSMs) so that the queue profile prediction algorithm knows the current location and speed of each vehicle. Furthermore, depending on the time in the signal cycle and current queue length when eco-driving vehicle arrives at the intersection, approaching vehicles downstream of the eco-driving vehicle may or may not stop based on current signal status and remaining timing of the signal phase. The details of calculating the start of the green window using the SPM can be found in~\cite{Zhen2018Traj}.

\vspace{-0.15cm} 
\section{Power Balance-Based HEV Model}
\begin{figure}[h!]
	\begin{center}
		\includegraphics[width=0.9\columnwidth]{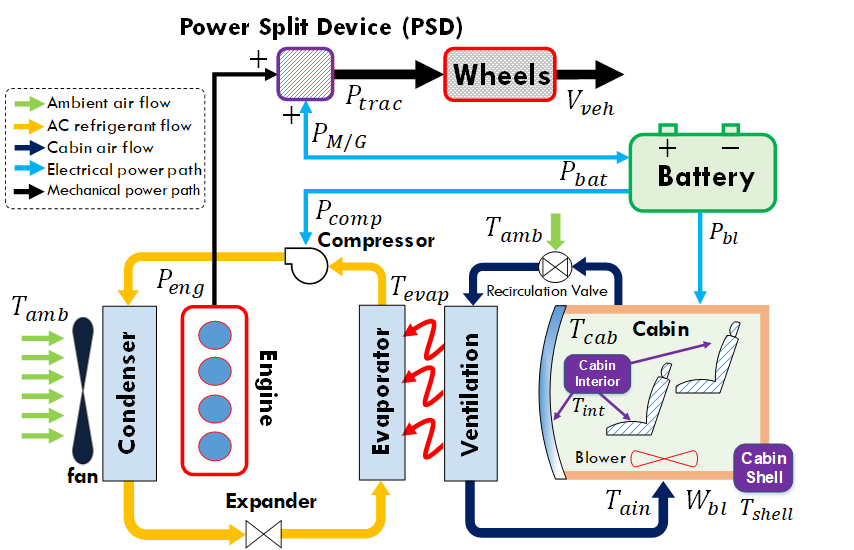} \vspace{-0.4cm}   
		\caption{Schematic of power-split HEV thermal and power loops.}\vspace{-0.85cm} 
		\label{fig:HEV_PS_AC_Schematic} 
	\end{center}
\end{figure}

\vspace{-0.1cm} 
The overall schematic of a power split HEV thermal and power loops is shown in Fig.~\ref{fig:HEV_PS_AC_Schematic}. The battery power ($P_{bat}$) is consumed to provide the demanded traction power, the A/C compressor power ($P_{comp}$), and HVAC blower power ($P_{bl}$). The A/C loop draws power from the battery to cool the cabin air temperature ($T_{cab}$). The traction power ($P_{trac}$) is provided via the power split device (PSD), which blends the engine output power ($P_{eng}$) and the electric propulsion from the motor ($P_{mot}$) and generator ($P_{gen}$). The total electric propulsion power is denoted by $P_{M/G}=P_{mot}+P_{gen}$. \vspace{-0.15cm}   

\subsection{Control-Oriented Model for A/C Load Optimization}\vspace{-0.1cm} 
A control-oriented A/C model was presented in our previous work~\cite{hwang2018}. This model predicts cabin air temperature and evaporator wall temperatures, and computes the power consumed by the A/C compressor and blower motors. The model has been verified against a high-fidelity physics-based model of an electrically driven A/C system, CoolSim\textsuperscript{\textregistered}, developed by the National Renewable Energy Lab (NREL)~\cite{kiss2013new}. For prediction in MPC, the following discrete-time A/C system model has been proposed~\cite{hwang2018}: \vspace{-0.2cm} 
\begin{gather}
\label{eqn:HVACModel_Tcab}
T_{cab}(k+1)=~~~~~~~~~~~~~~~~~~~~~~~~~~~~~~~~~~~~~~~~~~~~~~\\f_{T_{cab}}\big(T_{cab}(k),T_{int}(k),T_{shell}(k),T_{evap}(k),W_{bl}(k)\big)\nonumber\\
%
\label{eqn:HVACModel_Tevap}
T_{evap}(k+1)=f_{T_{evap}}\big(T_{evap}(k),T_{evap}^{s.p.}\big)~~~~~~~~~~~~~~~~
%
\end{gather} 
where $T_{cab}$, $T_{int}$, $T_{shell}$, and $T_{evap}$ represent the temperatures of the cabin air, the cabin interior (e.g., seats and panels), the cabin shell, and the evaporator wall, respectively (see Fig.~\ref{fig:HEV_PS_AC_Schematic}). The control inputs to the model are $W_{bl}$ (blower flow rate) and $T_{evap}^{s.p.}$ (evaporator wall temperature setpoint). The dynamics of $T_{bat}$ and $T_{evap}$ are captured through $f_{T_{cab}}$ and $f_{T_{evap}}$ nonlinear functions, respectively. See~\cite{hwang2018} for the details of the A/C system model (\ref{eqn:HVACModel_Tcab})-(\ref{eqn:HVACModel_Tevap}) and the identification and validation results.
~Furthermore, the power consumed by two major actuators of the A/C system, the compressor ($P_{comp}$) and the blower ($P_{bl}$), can be estimated as nonlinear functions of $f_{P_{comp}}(W_{bl},T_{amb},T_{evap})$ and $f_{P_{bl}}(W_{bl})$, respectively. See~\cite{hwang2018,Kelman11} for the details.
%
%

\vspace{-0.1cm} 
\subsection{Battery Power-Balance Model}\vspace{-0.1cm} 
%

\begin{figure*}[h!]
	\begin{center}
		\includegraphics[width=1.55\columnwidth]{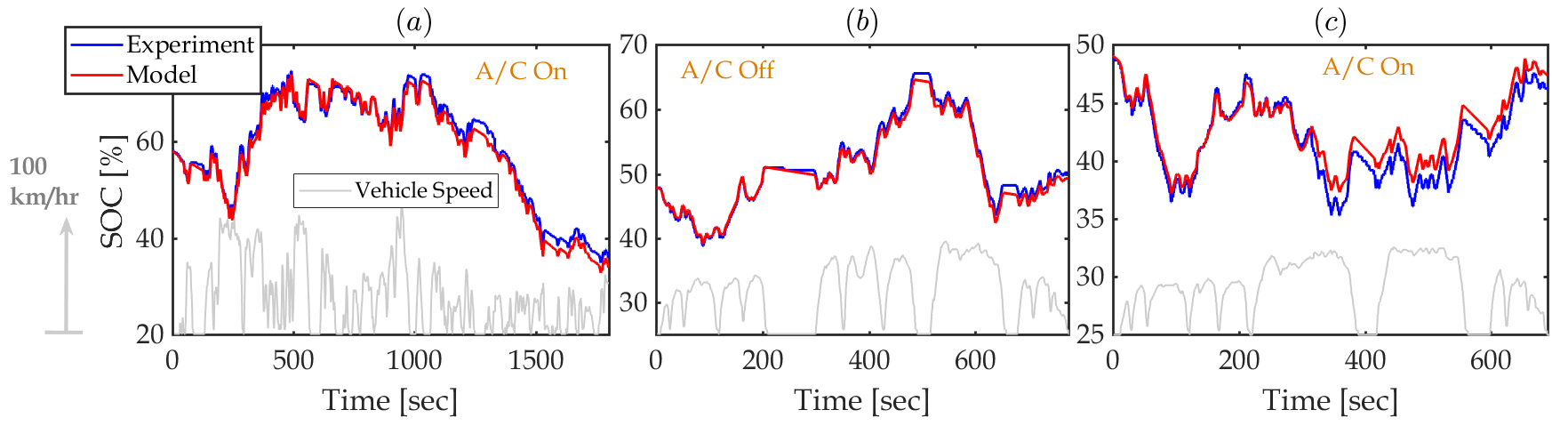} \vspace{-0.45cm}   
		\caption{The results of simplified battery SOC model validation. Experimental data collected from a Toyota Prius 2017 HEV at different A/C system operating settings. The driving cycle (vehicle speed) is shown in each sub-plot in gray.}\vspace{-0.75cm} 
		\label{fig:SOC_Model_Vrf} 
	\end{center}
\end{figure*}

In this section, a simplified phenomenological HEV powertrain model based on the mechanical and electrical power balance is developed, and experimentally validated. The main dynamics captured by the model are the battery state of charge ($SOC$), based on which the ratio between mechanical (engine) power and electrical (battery) power is determined by the powertrain controller. The proposed model, in addition to the traction power, captures the impact of the auxiliary A/C power load ($P_{A/C}=P_{comp}+P_{bl}$). Note that in our case $P_{trac}$ and $P_{A/C}$ are assumed to be the main power loads on the battery, and other auxiliary loads on the battery are neglected. The inputs to the model are the electric propulsion power ($P_{M/G}$), and the power consumed by the A/C system ($P_{A/C}$). The model output is the $SOC$ prediction. The $SOC$ prediction model has the following structure: \vspace{-0.25cm}   
%
\begin{gather}\label{eq:SOC_simple_model}
SOC(k+1)=SOC(k)+~~~~~~~~~~~~~~~~~~~~~~~~~~~~~~~\\
\begin{cases}
\xi_1 P_{M/G}(k)+\xi_2 P_{M/G}^2(k)+\xi_3 P_{M/G}(k)P_{A/C}(k)\\
~~~~~+\xi_4 P_{A/C}(k)+\xi_5 P_{A/C}^2(k)+\xi_6,~~~\text{if {A/C} is ON}\\
\xi_7 P_{M/G}(k)+\xi_8 P_{M/G}^2(k)+\xi_9,~~~~~~~~~\text{if {A/C} is OFF} 
\end{cases}\nonumber
\end{gather}
where sampling time is 1 $sec$, and $[\xi_1,\xi_2,\cdots,\xi_9]=-4.74\times10^{-5}$,$-4.11\times10^{-10}$,$6.17\times10^{-9}$,$-3.8\times10^{-5}$,$8.63\times10^{-9}$, $-0.03$,$-4.46\times10^{-5}$,$-4.84\times10^{-10}$,$-0.01]$ are the constant model parameters identified from experimental data collected from our test vehicle (2017 MY Prius). The proposed model employs a switching structure to represent different battery $SOC$ dynamic modes depending on whether the A/C system is on or off. Note that the model~(\ref{eq:SOC_simple_model}) accounts for the motor and generator efficiencies implicitly, as these efficiencies are being captured through the test data. The test data used for identification and validation of the model (\ref{eq:SOC_simple_model}) cover different testing scenarios with different A/C system operating settings and with different vehicle speed profiles. Fig.~\ref{fig:SOC_Model_Vrf} shows the $SOC$ model validation results. As can be observed from Fig.~\ref{fig:SOC_Model_Vrf}, the $SOC$ prediction error is less than $1\%$ on average ($max$ error$<3\%$). Thus, given the simple structure of the model and acceptable accuracy in predicting the battery state of charge, the model (\ref{eq:SOC_simple_model}) is used in the reminder of this paper to evaluate the HEV fuel economy, and design the rule-based and DP-based power split controllers.
%
%

\vspace{-0.1cm} 
\section{Sequential Optimization of Speed, Thermal Load, and Power Split} 
\begin{figure}[h!]
	\begin{center}
		\includegraphics[width=0.79\columnwidth]{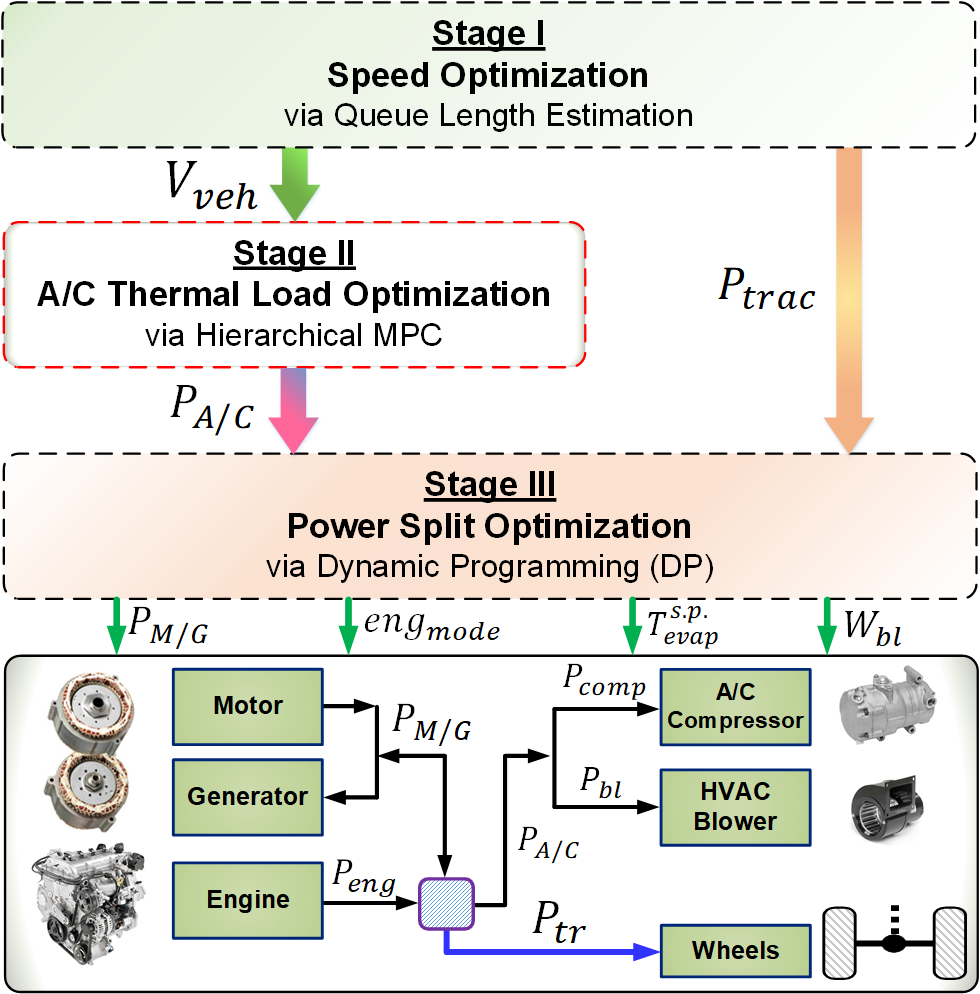} \vspace{-0.35cm}   
		\caption{Schematic of the proposed sequential HEV speed, thermal load, and power split optimization.}\vspace{-0.75cm} 
		\label{fig:HEV_Sequential_Optimization} 
	\end{center}
\end{figure}

\vspace{-0.2cm} 
In this section, the three stages of the proposed sequential optimization framework are described. As shown in Fig.~\ref{fig:HEV_Sequential_Optimization}, first, the vehicle speed optimization based on the predicted queuing dynamics is discussed. The outputs from the first optimization stage are the optimized vehicle speed and traction power trajectories, which are used as inputs to the subsequent optimization stages. In the second stage, the A/C load trajectory is optimized given the optimized vehicle speed profile derived in the first stage. The optimized A/C power trajectory from the second stage along with the optimized traction power from the first stage are used in the third optimization stage, where the optimal power split between mechanical engine power and battery electrical power, as well as the engine operating mode are determined. The details of each optimization stage are elaborated in the following subsections. \vspace{-0.2cm}  

\subsection{\textbf{Stage I}: Vehicle Speed Optimization}\vspace{-0.1cm}  
%
The green window gives the earliest and latest time instants that the eco-driving vehicle may arrive at the intersection, then the speed optimization algorithm generates an eco-friendly vehicle speed trajectory. The planning horizon of the trajectory starts from the time instant the eco-driving vehicle enters the communication range until it arrives at the intersection. In order to ensure a smooth speed trajectory and reduce the fuel consumption, a trigonometric speed profile from~\cite{barth2011dynamic} is used and defined as:\vspace{-0.15cm}
%
\begin{gather}\label{eq:traj_planning}
V_{veh}=
\begin{cases}
v_p-v_rcos(mt), t\in[0,t_p)\\
v_p-v_r\frac{m}{n}cos[n(t+\frac{\pi}{2n}-t_p)], t\in[t_p,t_q)\\
v_p+v_r\frac{m}{n}, t\in[t_q, t_{arr}),
\end{cases}
\end{gather}
where $v_p=d_{stop}/t_{arr},~v_r=v_p-v_0$, $V_{veh}$ is the vehicle speed, $v_0$ is the initial vehicle speed, $d_{stop}$ is the distance to stop bar, $t_{arr}$ is the time of arrival at the intersection given by ${t_{arr}=t_3+h}$, and $h$ is the saturation headway between two vehicles. At the time instant $t_p$, the speed of the eco-driving vehicle reaches the average speed $v_p$. After $t_q$, the vehicle speed does not change and the vehicle will cruise to the stop bar. The variables $m$ and $n$ are model parameters that determine the maximum acceleration, maximum deceleration and jerk. These parameters determine the shape of the trigonometric profile to minimize fuel consumption by reaching the cruise segment as soon as possible. See~\cite{barth2011dynamic} for more details. 

Based on different time instants that the eco-driving vehicle enters the communication range and corresponding queue status, the eco-driving vehicle may choose different speed profiles. Four types of speed profiles are identified: ``slow down'', ``speed up'', ``cruise" and ``stop". All ``slow down'', ``speed up'' and ``stop'' speed profiles are generated by the trigonometric profiles with different parameters, while the ``cruise'' speed profile maintains a constant speed to pass the intersection. In some previous studies, the ``stop'' speed profile is ignored because the eco-driving vehicle can always slow down to a very low speed and cross the intersection without a stop. However, a very low cruise speed may be disruptive to other traffic participants and cause frequent lane changing and cut-in behaviors. As a result, the minimum cruise speed is set to 70\% of the speed limit. If the eco-driving vehicle cannot maintain the minimum cruise speed, it will come to a complete stop.\vspace{-0.15cm} 

\subsection{\textbf{Stage II}: MPC-Based A/C Thermal Load Optimization}

The A/C thermal load of the HEV is optimized by implementing an H-MPC solution to optimize the cabin temperature setpoint ($T_{cab}^{s.p.}$), and, consequently, the compressor load while exploiting traffic and vehicle speed preview. We studied the sensitivity of the A/C system power consumption at different vehicle speeds in our previous work~\cite{hwang2018}, based on which the A/C efficiency map at different cabin temperature setpoints and vehicle speeds has been constructed. It was shown that the efficiency of the A/C system increases by approximately 30\% as the vehicle speed increases from 0 $m/s$ (stop condition) to 25 $m/s$. The impact of vehicle speed on the A/C system power consumption is even more pronounced if considering energy consumption normalized by the traveling distance. The A/C efficiency map facilitates incorporating traffic data into the A/C optimization problem. The main idea is to relax the requirement of constant cabin temperature setpoint tracking in the conventional A/C control system and allow a variable setpoint within the cabin temperature comfort zone, and then design a tracking MPC to follow the scheduled energy-optimal cabin temperature setpoint trajectory. Specifically, we are interested to optimize the load on the compressor by slightly varying $T_{cab}^{s.p.}$ to take advantage of the high efficiency range of the A/C system while vehicle is running at high speed. By exploiting the thermal storage of the cabin, additional thermal load is imposed when A/C is more efficient. This allows the A/C load to be reduced when it is less efficient without significant compromise of the passenger comfort. The idea of optimizing $T_{cab}^{s.p.}$ with respect to traffic preview is referred to as eco-cooling.
%

The proposed H-MPC for eco-cooling, shown in Fig.~\ref{fig:HEV_Sequential_Optimization_HMPC}, has two layers. The upper layer, scheduling layer, plans the optimal $T_{cab}^{s.p.}$ over a long prediction horizon to account for the relatively slow thermal dynamics of the cabin air temperature. In this paper, the uncertainties associated with long horizon vehicle speed predictions are not considered. However, we showed in~\cite{AminiCDC18} that thermal load optimization can be based on the average traffic flow speed information over a long horizon, see~\cite{AminiCDC18,sun2015integrating} for more details. The lower layer of the A/C controller, the piloting layer, is a short-horizon MPC designed to track the planned cabin air temperature setpoints scheduled by the upper layer MPC. 

\subsubsection{\textbf{Scheduling Layer of A/C MPC}}
\begin{figure}[h!]
	\begin{center}
		\includegraphics[width=0.7\columnwidth]{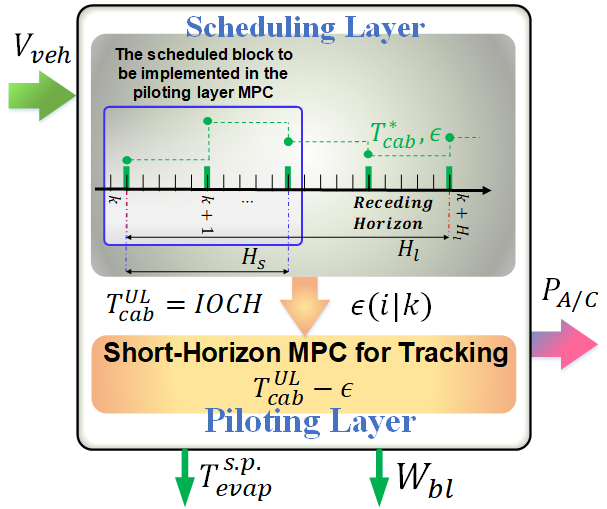} \vspace{-0.35cm}   
		\caption{Two-layer hierarchical A/C MPC. The calculated $T_{cab}-\epsilon(i|k)$ is sent from the scheduling layer to the piloting layer as the reference trajectory.}\vspace{-0.7cm} 
		\label{fig:HEV_Sequential_Optimization_HMPC} 
	\end{center}
\end{figure}

~The scheduling layer of the A/C MPC with traffic information incorporated is based on the solution of the following optimization problem:\vspace{-0.25cm}
\begin{equation} 
\small
\begin{aligned}\label{eqn:Intg_HVAC_IOCH}
&\argmin_{\substack{T_{evap}^{s.p.}(\cdot|k)\\W_{bl}(\cdot|k)\\\epsilon(\cdot|k)}} && \sum_{i=0}^{H_l} \Bigg\{ \begin{gathered} \overbrace{\frac{P_{comp}(i|k)}{\eta_{AC}(i|k)}+P_{bl}(i|k)}^{\ell_P}+\mathscr{B}\Big(\overbrace{\frac{\eta_{AC}(i)-1}{\epsilon(i|k)+\mathscr{D}}}^{\ell_{IOCH}}\Big)\\+\lambda\Big(\underbrace{T_{cab}^{s.p.}(i|k)-\frac{\sum_{i=0}^{H_l}T_{cab(i|k)}}{H_l+1}}_{\ell_{s.p.}}\Big)^2  \end{gathered}\Bigg\},\\
& \text{s.t.}
& & T_{cab}(i+1|k)=f_{T_{cab}}(i|k),~{i=0,\cdots,H_l},
\end{aligned}
\end{equation}
\begin{equation} 
\small
\begin{aligned}
&
& & T_{evap}(i+1|k)=f_{T_{evap}}(i|k),~{i=0,\cdots,H_l},\\
&
& &T_{cab}^{LL}\leq T_{cab}(i|k)\leq T_{cab}^{UL}-\epsilon(i|k),~{i=0,\cdots,H_l},\\
&
& &T_{evap}^{LL}\leq T_{evap}(i|k)\leq T_{evap}^{UL},~{i=0,\cdots,H_l},\\
&
& &0.05 \leq W_{bl}(i|k)\leq 0.15~kg/s,~{i=0,\cdots,H_l-1},\\
& 
& &3^oC\leq T_{evap}^{s.p.}(i|k)\leq 10^oC,~{i=0,\cdots,H_l-1},\\
& 
& &0^oC\leq \epsilon(i|k)\leq 3^oC,~{i=0,\cdots,H_l-1},\\
%
%
&
& & T_{cab}(0|k)=T_{cab}(k),~T_{evap}(0|k)=T_{evap}(k),\nonumber
\end{aligned}
\end{equation}
\normalsize
{where $(i|k)$ designates the prediction for the time instant $k+i$ made at the time instant $k$, and $H_l$ is the long prediction horizon.} The first two terms of the nonlinear MPC stage cost, $\ell_P$, represent the power consumption by the A/C system, where $\eta_{AC}$ is the A/C efficiency factor which is a function of the vehicle speed ($V_{veh}$). When $V_{veh}=0$, $\eta_{AC}=1$. Once $V_{veh}$ increases, $\eta_{AC}$ will increase to reflect the increase in the A/C system efficiency at higher speeds,~see~\cite{hwang2018}. The third term of the stage cost, $\ell_{IOCH}$, is incorporated for intelligent handling of the cabin temperature upper constraint ($T_{cab}^{U.L}-\epsilon(i|k)$), where $\epsilon(i|k)$ is a positive ``slack" optimization variable and $\mathscr{D}>0$ is a constant. The fourth term ($\ell_{s.p.}$) in the scheduling MPC stage cost is incorporated to reflect the objective of maintaining the average cabin temperature at the requested setpoint $T_{cab}^{s.p.}$. The parameters $T_{cab}^{UL}$ and $T_{cab}^{LL}$ are the upper and lower limits on the cabin temperature, which define the comfort zone inside the cabin. $T_{cab}^{LL}$ is set to 22$^oC$, while $T_{cab}^{UL}$ is variable. Furthermore, $T_{evap}^{UL}=12^oC$ and $T_{evap}^{LL}=0^oC$ are the upper and lower limits of the evaporator wall temperature. $\mathscr{B}$ and $\lambda$ are constant weighting factors to adjust the trade-off between A/C system energy consumption and cabin temperature setpoint tracking. The output of the scheduling A/C MPC is $T_{cab}^{UL}-\epsilon(i|k),~i=0,\cdots,H_p$, see Fig.~\ref{fig:HEV_Sequential_Optimization_HMPC}. \vspace{-0.45cm}

\subsubsection{\textbf{Piloting Layer of A/C MPC}}~The planned trajectory of $\tilde{T}_{cab}^{UL}(i|k)={T}_{cab}^{UL}-\epsilon(i|k),~i=0,\cdots,H_p$ from the scheduling layer is next passed on to the piloting layer MPC, which is designed to track the scheduled trajectories. The piloting layer MPC is defined based on the solution of the following optimization problem over a short horizon ($H_s$): \vspace{-0.15cm}
\begin{equation} 
\small
\begin{aligned}\label{eqn:piloting_HVAC_MPC}
& \argmin_{\substack{T_{evap}^{s.p.}(\cdot|k)\\W_{bl}(\cdot|k)}} & & \sum_{i=0}^{H_s}  \begin{gathered} \Bigg\{ \begin{gathered} P_{comp}(i|k)+P_{bl}(i|k)\\+w_{c}\Big(T_{cab}(i|k)-(T_{cab}^{UL}(i)-\epsilon(i|k))\Big)^2  \end{gathered}\Bigg\}\end{gathered}, \\
& \text{s.t.}
& & T_{cab}(i+1|k)=f_{T_{cab}}(i|k),~{i=0,\cdots,H_s},
\end{aligned}
\end{equation}
\begin{equation} 
\small
\begin{aligned}
&
& & T_{evap}(i+1|k)=f_{T_{evap}}(i|k),~{i=0,\cdots,H_s},\\
&
& &0.05 \leq W_{bl}(i|k)\leq 0.15~kg/s,~{i=0,\cdots,H_s-1},\\
& 
& &3^oC\leq T_{evap}^{s.p.}(i|k)\leq 10^oC,~{i=0,\cdots,H_s-1},\\
& 
& & T_{cab}(0|k)=T_{cab}(k),~T_{evap}(0|k)=T_{evap}(k).\nonumber
\end{aligned}
\end{equation}
\normalsize
As compared to the scheduling layer MPC (\ref{eqn:Intg_HVAC_IOCH}), the piloting layer MPC has fewer optimization variables and constraints, a less complicated cost function, and a shorter prediction horizon (e.g., $H_l=180~sec>H_s=30~sec$). \vspace{-0.2cm}

\subsection{\textbf{Stage III}: Power Split Optimization}
In the final optimization stage, the power split between engine mechanical and battery electrical powers is optimized by using Dynamic Programming (DP). For power split DP, the optimized A/C load calculated from hierarchical MPC (second stage), and optimized vehicle speed (traction power) from the first stage (as shown in Fig.~\ref{fig:HEV_Sequential_Optimization}) are used as known inputs, and the objective is to find the optimum battery power ($P_{bat}=P_{M/G}+P_{A/C}$) and engine operating mode to minimize the overall fuel consumption of the HEV, while enforcing the vehicle operating constraints. With a total travel time of $\mathcal{T}$ discretized by $K$ sampling instants, the most fuel efficient operation can be derived by minimizing the following cost function:\vspace{-0.22cm}
\begin{gather}
\label{eq:Task4_DPformulation}
{m_f(x,u,K)=\sum_{i=0}^K W_f\big(x(i),u(i)\big) +\Phi\big(x(K)\big)},\\
\text{s.t.}~~~~x(k+1)~\text{given by}~\text{Eq.~(\ref{eq:SOC_simple_model})},~x\in \mathcal{X},~u \in \mathcal{U},~~~~~
\end{gather}
where $x=SOC$ is the only state, the control input $u$ includes the engine operation mode and battery power as $u=[eng_{mode}, P_{bat}]$, and $\Phi(x(K))$ represents the terminal cost on $SOC$. Moreover, the following operation constraints in different operation modes should be satisfied:
\begin{itemize}
\item Engine Off:~$eng_{mode}=1$,~$P_{eng}=0,~\omega_e=0,~W_f=0$,
\item Engine On:~$eng_{mode}=2$,~$P_{eng}>0,~\omega_e\neq 0,~W_f=f(\omega_{eng},P_{eng})$,
\end{itemize}
%
where, $\omega_{eng}$ and $P_{eng}$ are the engine speed and engine output power, respectively. In the above, $f(\omega_{eng},P_{eng})$ designates the engine fuel consumption map as a function of the engine speed and power. This map has been developed from experimental data collected from the test vehicle (2017 MY Prius). When $eng_{mode}=2$, it is assumed that the engine operates on the optimal operation line, where fuel consumption is minimized for the given engine power. In this paper, we consider a rule-based power split controller with load leveling logic as the baseline, and we compare the proposed DP results against the rule-based controller. The rule-based controller is tuned to provide charge sustaining performance. The overall equivalent vehicle energy consumption is calculated to account for the net $SOC$ change by the end of the driving cycle, i.e., for the equivalent cost of the electrical energy consumption. \vspace{-0.1cm}
 
\section{Sequential Optimization Simulation Results}\vspace{-0.1cm}
\subsubsection{\textbf{Speed Optimization}}
\begin{figure}[h!]
 	\begin{center}
 		\includegraphics[width=0.85\columnwidth]{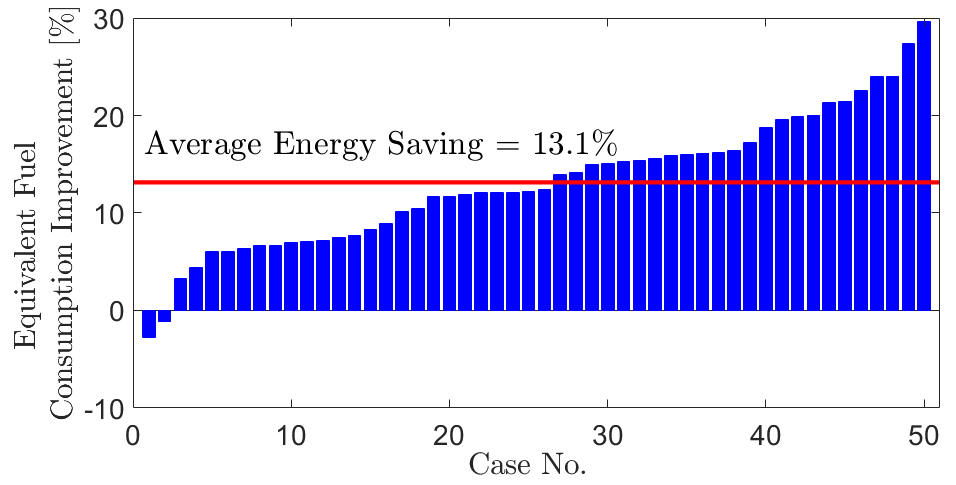} \vspace{-0.45cm}    		\caption{Energy saving results via speed optimization for 50 cases (vehicles) simulated over the route shown in Fig.~\ref{fig:AnnArbor_CAVs_Model}. The A/C controller tracks constant $T_{cab}^{s.p.}$=26$^oC$, and the powertrain controller is rule-based.}\vspace{-0.95cm} 
 		\label{fig:AllCases_SpeedOptz_EnergyComparison_Percent} 
 	\end{center}
\end{figure}

Fig.~\ref{fig:AllCases_SpeedOptz_EnergyComparison_Percent} shows the energy saving results achieved by the speed optimization at Stage I, while the A/C thermal load is not being optimized (namely constant $T_{cab}^{s.p.}$ is tracked), and the power split controller is rule-based. Fifty vehicles are simulated using different random speeds and go through the entire corridor shown in Fig.~\ref{fig:AnnArbor_CAVs_Model}. Compared with the non-optimized speed case, it can be seen from Fig.~\ref{fig:AllCases_SpeedOptz_EnergyComparison_Percent} that the overall energy consumption is reduced by 13.1\% on average, with the maximum reduction as high as 29.8\%. The variation in the energy saving results is caused by different strategies (e.g., ``speed up'' or ``cruise'') adopted by the vehicle with or without eco-driving. It can be seen that for two cases, the optimized speed profiles result in slightly higher energy consumption. It is possible that in these two cases, the speed trajectory planning is more conservative and stops more than vehicles with non-optimized speed trajectories, e.g.,  vehicle with non-optimized speed may pass the intersection when the signal turns to yellow, while the trajectory planning only considers green signal as the window to pass.

\begin{figure}[h!]
 	\begin{center}
 		\includegraphics[width=0.9\columnwidth]{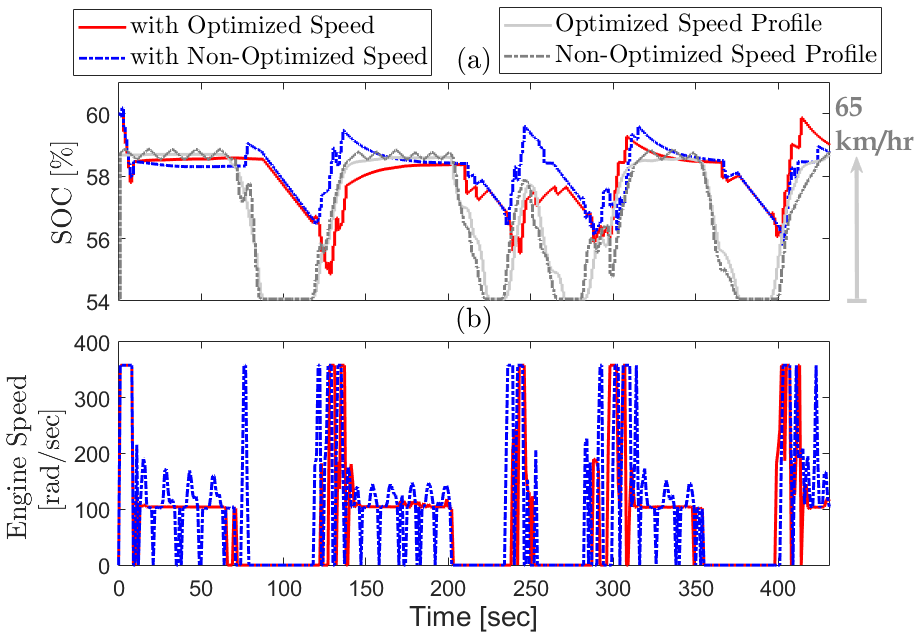} \vspace{-0.45cm}   
 		\caption{Results of vehicle speed optimization and its effects on vehicle powertrain performance: (a) battery $SOC$, and (b) engine speed. A/C controller tracks constant $T_{cab}^{s.p.}$. The powertrain controller is rule-based.}\vspace{-0.5cm} 
 		\label{fig:SpeedOptimization_SOCEng} 
 	\end{center}
 \end{figure}

Fig.~\ref{fig:SpeedOptimization_SOCEng}-a shows an example of the non-optimized speed trajectory (blue) versus the optimized one (gray). The response of $SOC$ and engine speed are also shown in Fig.~\ref{fig:SpeedOptimization_SOCEng}, when the A/C controller tracks constant $T_{cab}^{s.p.}$, and power split command is determined by the rule-based controller. It is observed that the speed planning algorithm generates smoother speed profiles, which consequently results in gentle use of the engine (Fig.~\ref{fig:SpeedOptimization_SOCEng}-b). Additionally, during vehicle stops, the engine is being used less often to charge the battery (Fig.~\ref{fig:SpeedOptimization_SOCEng}-a) while the battery supplies $P_{A/C}$. \vspace{-0.35cm}  
 
\subsubsection{\textbf{A/C Thermal Load Optimization}}
\begin{figure}[t!]
 	\begin{center}
 		\includegraphics[width=0.9\columnwidth]{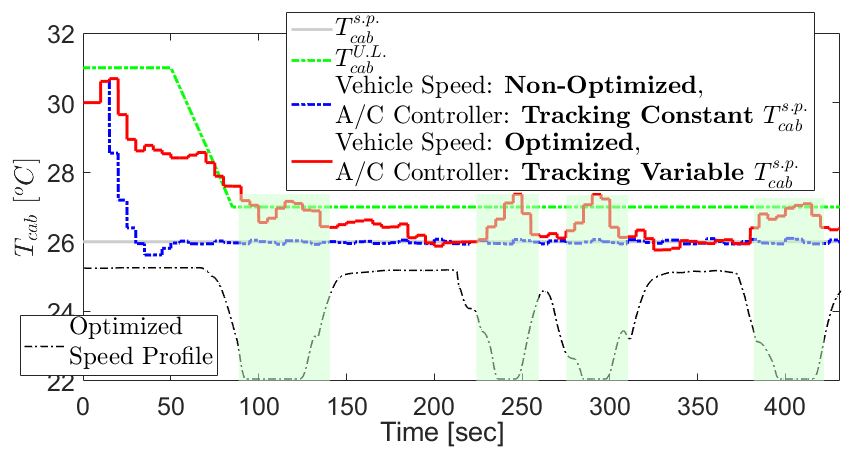} \vspace{-0.4cm}   
 		\caption{A/C power optimization via long-horizon MPC with IOCH implemented on CoolSim. For the first case (blue line), the speed is not optimized, and the A/C controller tracks constant $T_{cab}^{s.p.}$. For the second case (red), vehicle speed and A/C thermal load are optimized sequentially. Powertrain controller is rule-based in both cases.}\vspace{-0.95cm} 
 		\label{fig:ACoptimization_Tcab_Pac} 
 	\end{center}
 \end{figure}

The results of A/C thermal load optimization at Stage II of the proposed HEV optimization framework are shown in Fig.~\ref{fig:ACoptimization_Tcab_Pac}, where the cabin air temperature trajectories of the non-optimized case (tracking constant $T_{cab}^{s.p.}$) are compared with the hierarchical A/C MPC for eco-cooling. As highlighted in Fig.~\ref{fig:ACoptimization_Tcab_Pac}, the scheduling layer MPC of the A/C controller increases the cabin temperature setpoint by up to 1$^oC$ during vehicle stops to save the battery energy in low A/C efficiency regions. Moreover, during the first 80 $sec$ of the A/C operation when the cabin air temperature is still relatively higher than the target setpoint of 26$^oC$ (i.e., cool down period), the two-layer A/C MPC follows the upper bound of the cabin temperature comfort zone ($T_{cab}^{UL}$) to minimize the load on the A/C compressor. On the other hand, the conventional A/C controller, which is designed to track constant $T_{cab}^{s.p.}$, puts the effort to decrease the cabin temperature and ignores the inefficient speed-dependent periods of the A/C system operation. The modified results shown in Fig.~\ref{fig:ACoptimization_Tcab_Pac} are based on the optimized vehicle speed trajectory planned in Stage I, but the powertrain controller remains to be rule-based. By the end of the simulated driving cycle and after the cabin cool down period, an average cabin temperature of 26.2$^oC$ is resulted by using the conventional A/C controller for tracking constant $T_{cab}^{s.p.}$, while the average $T_{cab}=26.4^oC$ is observed from the proposed hierarchical A/C MPC. As it will be shown later, while the average cabin temperature at the steady state is slightly increased, the overall energy consumption with the optimized A/C thermal load is reduced by 2.4\% for the studied optimized speed profile. 
 
\subsubsection{\textbf{Power Split Optimization}} 
\begin{figure}[h!]
\begin{center}
 		\includegraphics[width=0.92\columnwidth]{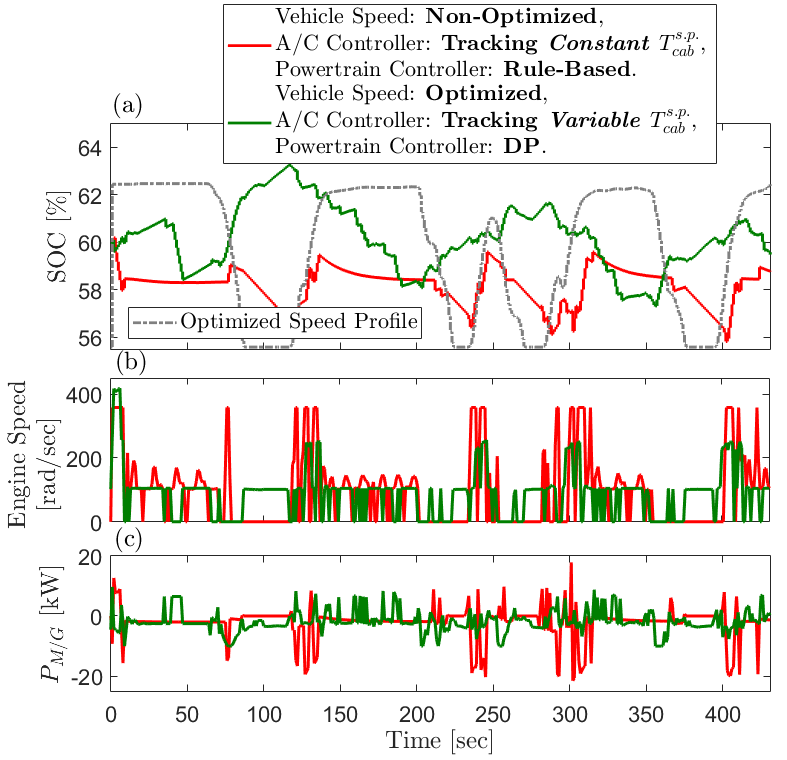} \vspace{-0.45cm}   
 		\caption{The results of sequential vehicle speed, thermal load, and powertrain optimization and its effects on vehicle powertrain performance compared to the baseline case: (a) battery $SOC$, (b) engine speed, and (c) battery electric propulsion.}\vspace{-0.65cm} 
 		\label{fig:DP_SpeedOptimization_SOCEng} 
\end{center}
\end{figure}

\begin{figure}[t!]
\begin{center}
 		\includegraphics[width=1.0\columnwidth]{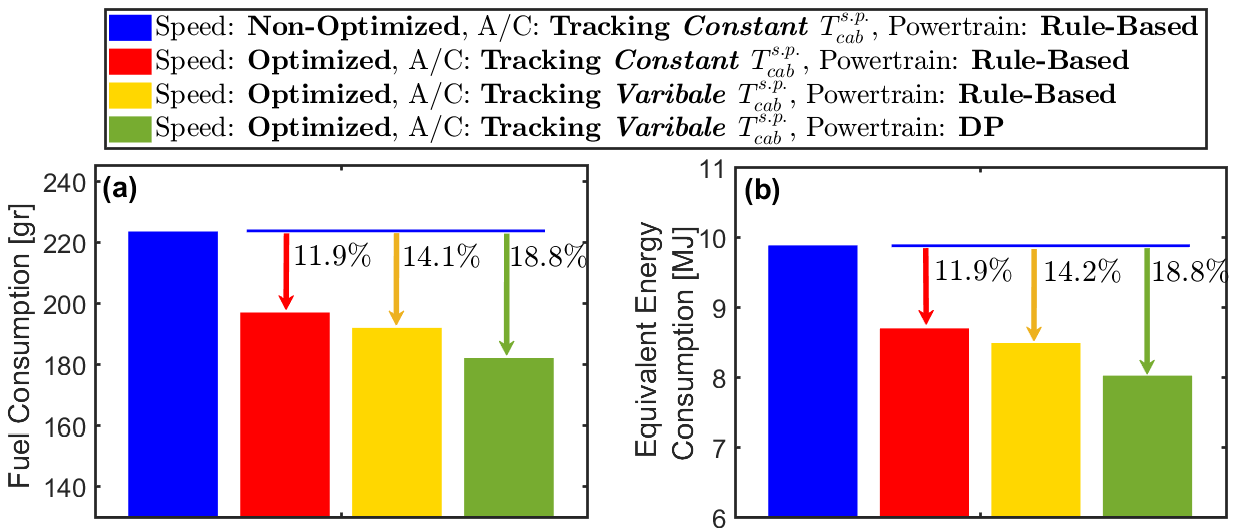} \vspace{-0.75cm}   
 		\caption{Results of (a) fuel and (b) energy saving after applying each stage of the proposed sequential optimization technique.}\vspace{-1.05cm} 
 		\label{fig:SpeedOptimization_FuelEnergy} 
\end{center}
\end{figure}

Figs.~\ref{fig:DP_SpeedOptimization_SOCEng} and \ref{fig:SpeedOptimization_FuelEnergy} depict the results of applying the third and final optimization stage for HEV power split. It is shown in Figs.~\ref{fig:DP_SpeedOptimization_SOCEng} that the proposed sequential optimization framework results in entirely different powertrain operation. For the baseline case, the battery is being used during vehicle stops for the A/C system operation and the engine is off (the battery is not being charged). On the other hand, the optimized case shows that the battery is being charged during vehicle stops, while the thermal load on the battery is reduced by the hierarchical A/C MPC. The engine is being used more often for the optimized case while it runs at relatively lower (optimal) and constant speed; consequently the overall fuel consumption is lower compared to the baseline case by the end of the driving cycle. Compared to the optimized case, the baseline case utilizes the engine more aggressively during the vehicle accelerations with non-smooth engine speed trajectories. Additionally, according to Fig.~\ref{fig:DP_SpeedOptimization_SOCEng}-c, the optimized case uses the electrical power for propulsion more often using the energy saved in the battery during the stops. This results in less usage of the engine for traction, thereby lowering fuel and energy consumption.  

Fig.~\ref{fig:SpeedOptimization_FuelEnergy} summarizes the overall fuel and energy consumption of the HEV after applying each optimization stage, and compares the results with the baseline case for the speed profile shown in Fig.~\ref{fig:SpeedOptimization_SOCEng}-a. It can be observed that the major saving in energy consumption is attributed to the speed optimization (11.9\%). Eco-cooling improves the energy saving from 11.9\% to 14.2\%, and by applying the final power split optimization, cumulative energy saving of 18.8\% is achieved.\vspace{-0.25cm}

%
\section{Summary and Conclusions}\label{sec:5}
This paper presented a multi-stage sequential optimization solution for thermal and power management of HEVs to aggressively reduce energy losses using predictive traffic information. The main objective is to first optimize the traction and thermal powers, then calculate the optimum power split ratio between the mechanical and electrical propulsion systems of the HEV. To this end, an eco-driving vehicle speed trajectory planning in a congested urban environment is implemented at the first optimization stage according to the information collected from both traffic signals and vehicle queue at the intersections. Given the A/C system speed sensitivity, in the second stage, and with respect to the planned speed trajectory from the first stage, a hierarchical A/C MPC is used to find the optimal thermal load trajectory based on a control-oriented model of the thermal system. In the third stage, the optimal traction and thermal power trajectories are used to inform the optimal HEV powertrain operation by applying Dynamic Programming (DP) to a simplified and experimentally validated power-balance model of the HEV. Compared to a baseline HEV with non-optimized speed and thermal loads and a rule-based powertrain controller, simulation results for a real-world corridor of six intersections show that the eco-driving speed trajectory planning from the first optimization stage can significantly reduce the CAVs energy consumption by up to 29.8\%, with an average saving of 13.1\% for fifty vehicles studied over the same corridor. Additionally, for one selected vehicle, while up to 11.9\% energy saving is observed from speed optimization stage, further energy savings of up to 14.2\% and 18.8\% can be achieved after applying the second (eco-cooling) and third (power split) stages of the proposed sequential optimization framework, respectively. 
\vspace{-0.25cm}


\bibliographystyle{unsrt} 
\bibliography{ACC2018Ref.bib} \vspace{-0.25cm}



\end{document}